\documentclass[prc,twocolumn,nofootinbib,showpacs,floatfix,byrevtex,
letterpaper]{revtex4}

\usepackage{graphicx,slashbox}
\usepackage{amsmath,amssymb}
\usepackage[textheight=23cm,textwidth=18cm,top=2.5cm]{geometry}
\graphicspath{{./figures/}}
\newcommand{\eq}[1]{\begin{equation} #1 \end{equation}}
\newcommand{\expval}[1]{\langle #1 \rangle}

\newcommand{\ket}[1]{|#1\rangle}
\newcommand{\Dcm}{D_{\mathrm{c.m.}}}

\newcommand{\EKcm}{E_{\mathrm{K}}}

\begin{document}


\title{Fission modes of $^{256}$Fm and $^{258}$Fm in a microscopic approach}

\author{L. Bonneau}

\affiliation{Theoretical Division, Los Alamos National Laboratory, 
Los Alamos, NM 87545, USA}

\date{\today}

\begin{abstract}
A static microscopic study of potential-energy surfaces within the 
Skyrme-Hartree-Fock-plus-BCS model is carried out for the 
$^{256}$Fm and $^{258}$Fm isotopes with the goal of deducing some properties 
of spontaneous fission. The calculated fission modes are found to be 
in agreement with the experimentaly observed asymmetric-to-symmetric 
transition in the fragment-mass distributions and with the high- and 
low-total-kinetic-energy modes experimentally observed in $^{258}$Fm. 
Most of the results are similar to those obtained in 
macroscopic-microscopic models as well as in recent 
Hartree-Fock-Bogolyubov calculations with the Gogny interaction, 
with a few differences in their interpretations. In particular an alternative 
explanation is proposed for the low-energy fission mode of $^{258}$Fm.
\end{abstract}


\maketitle

\section{Introduction}

Before the work of Brandt and collaborators~\cite{Brandt} in 1963, very 
little was experimentally known about detailed spontaneous-fission 
properties of isotopes other than $^{252}$Cf. These authors measured the 
mass-yield and kinetic-energy distributions for the spontaneous fission 
of $^{254}$Fm and obtained a well-marked asymmetric mass-yield curve. 
Later on, Balagna \textit{et al.}~\cite{Balagna} and John \textit{et al.}%
~\cite{John} showed independently that the experimental fragment-mass 
distribution for the spontaneous fission of $^{257}$Fm is essentially 
symmetric, with a very slight dip in a broad peak centered 
at the fragment mass 127--128. John and collaborators~\cite{John} also 
investigated the thermal-neutron-induced fission of $^{257}$Fm and obtained 
a sharper symmetric pattern. Based on the results from the Argonne group 
published the following year~\cite {Flynn} and showing the asymmetric 
character of the spontaneous fission of $^{256}$Fm, they concluded that 
``symmetric mass division in low-energy fission of heavy actinides appears 
abruptly at $^{257}$Fm''. This was confirmed later in the measurements 
by Hulet and collaborators~\cite{Hulet_Fm259,Hoffman_systematics,
Hulet_Fm258_PRL1986} who found very narrow symmetric mass distributions 
in the spontaneous fission of $^{258}$Fm and $^{259}$Fm.

Another remarkable spontaneous-fission property that rapidly changes 
along the Fm isotopic chain is the fragment kinetic-energy 
distribution: it is very well reproduced by a single Gaussian 
for $A\leqslant 256$ with an average total kinetic energy $\overline{\rm TKE}$ 
that follows the Viola systematics~\cite{Hoffman_systematics,Viola1985}, 
whereas heavier isotopes have a much higher $\overline{\rm TKE}$-value. 
This is particularly the case for $^{258}$Fm where Hulet \textit{et 
al.}~\cite{Hulet_Fm258_PRC1989} found that the total-kinetic-energy 
distribution has a non-Gaussian shape which can be unfolded into two
Gaussians. These authors named this behavior 
``bimodal fission'' and showed that it corresponds to two different fission 
modes: one is a high-energy mode ($\overline{\rm TKE}=230$ MeV) associated 
with a narrow symmetric mass distribution, the other one is a low-energy 
($\overline{\rm TKE}=205$ MeV) form of fission with a much broader (still 
symmetric) mass distribution, which even reverts to asymmetric when 
spontaneous-fission events associated with lower energies ($\rm TKE\leqslant 
200$ MeV) are selected. Interestingly both modes have about the same abundance.

From the theory side, a number of authors investigated these two spontaneous-%
fission properties in Fm isotopes. On the one hand, Lustig, Maruhn and Greiner%
~\cite{Lustig} studied the transition in the mass-distribution patterns. They 
calculated the mass-yield curves of even Fm isotopes from $A=254$ to $A=260$ 
and reproduced the observed transition. On the other hand, 
several groups investigated the bimodal fission of $^{258}$Fm, from the 
characteristics of its potential-energy surface. Four of them made use 
of a macroscopic-microscopic model, relying on a liquid-drop 
contribution to the binding energy (with different nuclear surface 
parametrizations) and the Strutinsky method with various single-particle 
potentials to calculate the microscopic correction to the macroscopic energy%
~\cite{Brosa_ZPA1986,Moller_1987-1989,Moller_Nature,Pashkevich,Cwiok}. Even 
though all of them did not agree on the height of the outer saddle point 
relative to the ground state of the fissioning nucleus, they all 
obtained two fission valleys leading to two different families of prescission 
shapes, namely elongated slightly asymmetrical configurations, and 
compact symmetrical configurations corresponding to two nearly spherical 
$^{129}$Sn nascent fragments. They were dubbed respectively as 
the ``old path'' and the ``new path'' in 
Refs.~\cite{Moller_1987-1989,Moller_Nature}. The existence of these two 
fission valleys is in agreement with the experimental observation of 
two energy modes. Pashkevich~\cite{Pashkevich} even showed that the 
difference between the Coulomb interaction energy between the nascent 
fragments estimated in both valleys for different neck radii (about 
30 MeV) was reasonably close to the TKE difference between the 
fission modes (25 MeV).

More recently, two groups of authors performed calculations of potential-%
energy surfaces, covering the region of saddle points and beyond, within 
microscopic self-consistent mean-field approaches including residual pairing 
correlations and using density-dependent phenomenological effective 
nucleon-nucleon interactions. On the one hand, within the 
Hartree-Fock-Bogolyubov (HFB) approximation with the D1S Gogny interaction, 
Warda and collaborators~\cite{Warda2002,Warda2003,Warda2004} carried out an 
extensive study addressing both spontaneous-fission properties discussed in 
the first two paragraphs. However some of their arguments do not seem to be 
consistent with their results, as will be shown in Sect.~3. On the other 
hand, Staszczak and coworkers~\cite{Staszczak} implemented the Hartree-Fock-%
plus-BCS (HFBCS) approximation, with the SLy4 Skyrme interaction in the mean-%
field channel and the seniority (constant-$G$) force in the pairing channel.  
They also obtained a reflection-asymmetric path and a symmetric one in 
their fission-barrier calculations for the even Fm isotopes from $A=242$ 
to $A=264$, but their claimed overall qualitative agreement with existing 
experimental data does not seem consistent with the results presented. 

Finally it is worth mentioning that Asano and coworkers~\cite{Asano} recently 
performed dynamical calculations of fragment kinetic-energy and mass 
distributions, based on the multi-dimensional Langevin equation and a 
macroscopic-microscopic potential-energy surface, for the $^{264}$Fm isotope 
at a compound nucleus excitation energy of 10~MeV. They also calculated the 
mass distributions of $^{256}$Fm and $^{258}$Fm at the same excitation energy 
and obtained a good agreement with the experimental data for the 
thermal-neutron-induced fission of $^{255}$Fm~\cite{Ragaini} and $^{257}$Fm%
~\cite{John}.

Within the HFBCS microscopic approach, the present study aims at clarifying 
several points related to the above spontaneous-fission properties of 
$^{256}$Fm and $^{258}$Fm from a careful, static study of their 
potential-energy surfaces, extending the brief discussion in 
Ref.~\cite{Cadarache_PES}. After a short description of the 
HFBCS model and the method used to explore the 
energy surface, I will present the results in Sect.~3 and 
discuss them in Sect.~4. Finally I will draw the main conclusions in~Sect.~5.

\section{Model and method for exploring the potential-energy surface}

The microscopic approach followed in this study has already been 
applied to calculations of actinide fission barriers and presented in 
detail in Ref.~\cite{papier_fission_EPJA}. It is based on the 
Hartree-Fock-plus-BCS approximation implemented with a 
density-dependent nucleon-nucleon effective 
interaction. Such an approach is self-consistent owing to 
both the Hartree-Fock approximation itself and the density-dependence 
of the effective interaction. However, since very large elongations are 
involved up to and beyond scission (defined for example as in the recent 
works~\cite{Goutte_U238} and \cite{Cadarache_scission_spin}), the 
approximate correction term for the zero-point rotational motion added in 
Ref.~\cite{papier_fission_EPJA} does not seem to be well suited for such 
deformations and is not taken into account. As for the spurious 
contribution from center-of-mass vibrations to the energy, thoroughly analyzed 
by Bender \textit{et al.}\ in Ref.~\cite{Bender_COMcorrection}, only the 
traditional one-body contribution $\EKcm^{(1)}$ (leading simply to a 
renormalization factor $1-1/A$ in the kinetic energy) is considered 
since the SkM* force parameters were fitted within this framework. As we can 
see in Tab.~\ref{EKcm}, adding the two-body contribution $\EKcm^{(2)}$ 
(perturbatively calculated) to the Hamiltonian changes the deformation 
energy with respect to its ground-state value by about 1 MeV. 
\begin{table}[h]
\caption{One-body $\EKcm^{(1)}$ and two-body $\EKcm^{(2)}$ contributions to 
the center-of-mass kinetic energy in MeV per nucleon at the ground state 
($Q_{20}=32.5$~b) and far beyond the outer saddle point in the fission valley 
($Q_{20}\approx400$~b) of $^{256}$Fm (see Sect.~3 for details).\label{EKcm}}
\begin{center}
\begin{tabular}{ccccccccc}
\hline
\hline
 & $Q_{20}$ (b) && $\EKcm^{(1)}/A$ && $\EKcm^{(2)}/A$ && total
& \\
\hline
 & 32.5 && 18.72 && $-13.21$ && 5.51 & \\
 & 400.7 && 18.34 && $-$12.16 && 6.18 & \\
\hline
\hline
\end{tabular}
\end{center}
\end{table}
As long as we deal with one-cluster nuclear shapes, this effect can be 
considered to be small. However, the center-of-mass correction is expected 
to be important at and beyond scission, and a tractable correction appropriate 
for any configuration ranging from slightly deformed to well-separated 
fragment shapes has not yet been proposed. 

To illustrate this, let us consider a configuration with two identical, well-%
separated fragments. On the one hand, we would expect the energy of the 
total system to be
\eq{
\label{E_intuitive}
E=E_1+E_2+E_{\rm C}^{(\rm int)}\:,\\
}
where $E_{\rm C}^{(\rm int)}$ is the Coulomb interaction energy between 
the two fragments (which expression, not needed here, will be given after 
this discussion). In Eq.~(\ref{E_intuitive}), $E_i$ denotes 
the energy of fragment $i$ obtained by subtracting the one- and two-body 
parts of the center-of-mass kinetic energy of fragment $i$ from the expectation 
value $\expval{\hat{H}}(A/2)$ of the corresponding Hamiltonian:
\eq{
E_i=\expval{\hat{H}}(A/2)-\biggl(\frac{\EKcm^{(1)}(A/2)}{A/2}+
\frac{\EKcm^{(2)}(A/2)}{A/2}\biggr)\:.
}
Using the following estimates for $\EKcm^{(1)}/A$ and $\EKcm^{(2)}/A$
assumed to be independent of $A$ and the deformation coordinate
(expressed as the quadrupole moment $Q_{20}$) and 
deduced from Tab.~\ref{EKcm}:
\begin{gather}
\EKcm^{(1)}/A\approx 18.5\mbox{ MeV} \\
\EKcm^{(2)}/A\approx -0.7\EKcm^{(1)}/A\:,
\end{gather}
we arrive at
\eq{
E\approx 2\,\expval{\hat{H}}(A/2)+E_{\rm C}^{(\rm int)}-11\mbox{ MeV.}
}
On the other hand, a common application of the center-of-mass correction 
(even including the two-body part) leads to
\eq{
E'=\expval{\hat{H}}(A)-\biggl(\frac{\EKcm^{(1)}(A)}{A}+
\frac{\EKcm^{(2)}(A)}{A}\biggr)\:,
}
with $\expval{\hat{H}}(A) \approx 2\,\expval{\hat{H}}(A/2)+
E_{\rm C}^{(\rm int)}$; hence
\eq{
E'\approx 2\,\expval{\hat{H}}(A/2)+E_{\rm C}^{(\rm int)}-5.5\mbox{ MeV.}
}
The difference between $E$, the intuitively correct result, and $E'$ 
amounts to about 5 to 6 MeV in absolute value. In this work, where the 
two-body part of the center-of-mass correction is not taken into 
account, the energy actually calculated is
\begin{align}
E'' & =\expval{\hat{H}}(A)-\frac{\EKcm^{(1)}(A)}{A} \nonumber \\
& \approx 2\,\expval{\hat{H}}(A/2)+E_{\rm C}^{(\rm int)}-18.5\mbox{ MeV,}
\end{align}
which underestimates $E$ by about 7 to 8 MeV. That means that, in the absence 
of a correct treatment of the spurious center-of-mass motion, the approximate 
corrected energy calculated here is off by at least 5 MeV at and beyond 
scission, even if the two-body contribution is taken into account. 
Nevertheless for the purpose of the present study this 
is not a serious limitation since it does not affect the 
existence of stable solutions at a given elongation.

Let us now be precise about how the Coulomb energy is calculated
in this work and what expression of $E_{\rm C}^{(\rm int)}$ is 
used to estimate the fission-fragment total kinetic energy in Sect.~4. 
Firstly the direct Coulomb energy $E_{\rm C}^{(\rm dir)}$ of the 
fissioning nucleus
\begin{align}
E_{\rm C}^{(\rm dir)} &=\frac{e^2}{2}\int d^3\mathbf{r}
\int d^3\mathbf{r}'\:\frac{\rho_p(\mathbf{r})\,\rho_p(\mathbf{r}')}{
|\mathbf{r}-\mathbf{r}'|} \\
\label{well_behaved_Ecouldir}
&=\frac{e^2}{2}\int d^3\mathbf{r}\int d^3\mathbf{r}\,' \left|
\mathbf{r}-\mathbf{r}\,'\right| \Delta\rho_p(\mathbf{r}\,')\:,
\end{align}
where $e^2\approx 1.44\mbox{ MeV\,fm}$, $\rho_p(\mathbf{r})$ is the (local) 
proton density and $\Delta$ is the Laplacian operator, is computed 
exactly by numerical integration of the well-behaved 
expression~(\ref{well_behaved_Ecouldir}) as proposed by Vautherin%
~\cite{Vautherin}. Then the exchange part $E_{\rm C}^{(\rm exch)}$ is 
calculated by combining the Slater approximation~\cite{Slater} with a 
kind of local density approximation, which leads to the expression%
~\cite{Negele_Vautherin,Gombas}
\begin{align}
E_{\rm C}^{(\rm exch)} &=-\frac{e^2}{2}\int d^3\mathbf{r}
\int d^3\mathbf{r}'\frac{\bigl|\rho_p(\mathbf{r},\mathbf{r}')\bigr|^2}{
|\mathbf{r}-\mathbf{r}'|} \\
\label{Ecoul_exch_Slater}
&\approx-\frac{3\,e^2}{4}\biggl(\frac{3}{\pi}\biggr)^{1/3}
\int d^3\mathbf{r}\:\bigl[\rho_p(\mathbf{r})\bigr]^{4/3}\:,
\end{align}
where $\rho_p(\mathbf{r},\mathbf{r}')$ is the nonlocal 
proton density (off-diagonal matrix elements of the one-body density 
operator). The two densities $\rho_p(\mathbf{r})$ and 
$\rho_p(\mathbf{r},\mathbf{r}')$, including the BCS occupation probabilities 
$v_i^{\,2}$ as in Ref.~\cite{Vautherin}, take the form
\begin{gather}
\label{local_rho_p}
\rho_p(\mathbf{r})=2\,\sum_{i>0}v_i^{\,2}\sum_{\sigma}
\bigl|\varphi_i^{(\sigma)}(\mathbf{r})\bigr|^2 \\
\label{non_local_rho_p}
\rho_p(\mathbf{r},\mathbf{r}')=2\,\sum_{i>0}v_i^{\,2}\sum_{\sigma,\sigma'}
\bigl(\varphi_i^{(\sigma)}(\mathbf{r})\bigr)^*
\varphi_i^{(\sigma')}(\mathbf{r'})\:,
\end{gather}
where the sums run over all the pairs of time-reversed conjugate states and 
$\varphi_i^{(\sigma)}(\mathbf{r})$ stands for the component of spin $\sigma$ 
of the single-particle wave function associated with the single-particle 
state $\ket{i}$. The Coulomb interaction energy between two complementary 
pieces $\mathcal{V}_1$ and $\mathcal{V}_2$ of the fissioning nucleus
is defined by
\eq{
E_{\rm C}^{(\rm int)}=e^2\int_{\mathcal{V}_{\rm 1}}\!\!\!
d^3\mathbf{r_1}\int_{\mathcal{V}_{\rm 2}}\!\!\! d^3\mathbf{r_2}
\:\frac{\rho_p(\mathbf{r_1})\,\rho_p(\mathbf{r_2})-
\bigl|\rho_p(\mathbf{r_1},\mathbf{r_2})\bigr|^2}{|\mathbf{r_1}-\mathbf{r_2}|}
}
and can be decomposed into direct and exchange parts 
$E_{\rm C}^{(\rm int\:dir)}$ and $E_{\rm C}^{(\rm int\:exch)}$, respectively:
\eq{
E_{\rm C}^{(\rm int)}=E_{\rm C}^{(\rm int\:dir)}+
E_{\rm C}^{(\rm int\:exch)}\:,
}
with
\begin{gather}
E_{\rm C}^{(\rm int\:dir)}=e^2\int_{\mathcal{V}_{\rm 1}}\!\!\! 
d^3\mathbf{r_1}\:\rho_p(\mathbf{r_1})\int_{\mathcal{V}_{\rm 2}}\!\!\! 
d^3\mathbf{r_2}\:\frac{\rho_p(\mathbf{r_2})}{|\mathbf{r_1}-\mathbf{r_2}|} \\
\label{Ecoul_int_exch}
E_{\rm C}^{(\rm int\:exch)}=-\,e^2\int_{\mathcal{V}_{\rm 1}}\!\!\! 
d^3\mathbf{r_1}\int_{\mathcal{V}_{\rm 2}}\!\!\! d^3\mathbf{r_2}
\frac{\bigl|\rho_p(\mathbf{r_1},\mathbf{r_2})\bigr|^2}{
|\mathbf{r_1}-\mathbf{r_2}|}\:.
\end{gather}
Eqs.~(\ref{non_local_rho_p}) and (\ref{Ecoul_int_exch}) show that the 
exchange contribution vanishes when the single-%
particle wave functions are localized in either subset $\mathcal{V}_{\rm 1}$ 
or $\mathcal{V}_{\rm 2}$, which is the case in separated-fragment 
configurations (at and beyond scission), but not before the neck ruptures. 
Since the exchange part of the Coulomb energy 
is calculated with the approximate expression~(\ref{Ecoul_exch_Slater}), 
it gives no contribution to the Coulomb interaction energy. Indeed 
$E_{\rm C}^{(\rm exch)}$ can be written as
\eq{
E_{\rm C}^{(\rm exch)}=E_{\rm C}^{({\rm exch}\,1)}+
E_{\rm C}^{({\rm exch}\,2)}\:,
}
where $E_{\rm C}^{({\rm exch}\,i)}$ is given by
\eq{
E_{\rm C}^{({\rm exch}\,i)}=
-\frac{3\,e^2}{4}\biggl(\frac{3}{\pi}\biggr)^{1/3}
\int_{\mathcal{V}_{i}}\!\!\!  d^3\mathbf{r}\:\bigl[
\rho_p(\mathbf{r})\bigr]^{4/3}
}
and is interpreted as the exchange 
term of the Coulomb self-energy of the subset $\mathcal{V}_{i}$. In 
consequence, $E_{\rm C}^{(\rm int)}$ is overestimated when its 
exchange contribution is calculated with the 
approximation~(\ref{Ecoul_exch_Slater}), but by less and less 
as we approach scission.

In practice the phenomenological Skyrme SkM* interaction in the mean-field 
channel and the seniority force in the pairing channel are 
chosen, with the same set of parameters as in 
Ref.~\cite{papier_fission_EPJA}. The Hartree-Fock equations are solved 
by expansion of the single-particle states in a cylindrical harmonic-%
oscillator basis, which needs to be appropriately truncated for 
practical calculations. Appropriately means here that, given a truncation 
prescription and a maximal size parameter $N_0$ (see, e.g, 
Ref.~\cite{FQKV}), 
the basis parameters for a given deformation should be chosen so as to 
minimize the energy. Following Ref.~\cite{FQKV}, I introduce a deformation 
parameter $q=\omega_{\bot}/\omega_z$ and the spherical-equivalent
harmonic-oscillator constant $b=\sqrt{m\,\omega/\hbar}$, 
where $\omega$ is related to the oscillator frequencies 
$\omega_z$ (in the $z$ direction) and $\omega_{\bot}$ (in the plane
perpendicular to the $z$-axis) through $\omega^3=\omega_z\,\omega_{\bot}^2$. 
In particular 17 oscillator shells ($N_0=16$) are included throughout 
this work unless otherwise specified, as this basis size was shown 
in Ref.~\cite{papier_fission_EPJA} to be sufficient to within about 
0.1~MeV for the relative energies (with respect to the ground state) 
up to the outer saddle point of $^{252}$Cf.

Given the variational character of the HFBCS approximation and the difficulty 
of computing energy on a mesh in an $N$-dimensional deformation space 
with $N>2$, I~resort to implementing the method described in 
Ref.~\cite{papier_fission_PRC} where it was applied to calculations of
fission paths for $^{70}$Se. Assuming axial symmetry, a limited number
of shape degrees of freedom are retained, namely the elongation
of the fissioning system expressed as the 
axial quadrupole moment $Q_{20}$ or the center-of-mass distance $\Dcm$ 
between the (pre-)fragments, the mass asymmetry through either the axial 
octupole moment $Q_{30}$ or the heavy (pre-)fragment mass $A_H$, and the 
neck coordinate $Q_N$ (introduced by Berger \textit{et al.}%
~\cite{Berger_QN} and used by Warda \textit{et al.}~\cite{Warda2002}). 
Under this assumption, the center of mass of the fissioning nucleus, 
located on the symmetry axis chosen to be the $z$-axis, is fixed at 
the origin of the reference frame by adding to the Hamiltonian a constraint 
on the expectation value of $z$. The definitions of the retained shape 
coordinates can be found in Ref.~\cite{papier_fission_PRC}, except for~$A_H$
\eq{
A_H=\mathrm{max} \{A_{\rm right},A_{\rm left}\}
}
where the mass of the right and left fragments are respectively defined by
\begin{gather}
\label{Aright}
A_{\rm right}=\int_{z\geqslant z_{\rm
neck}}d^3\mathbf{r}\,\rho(\mathbf{r}) \\
\label{Aleft}
A_{\rm left}=A-A_{\rm right}\:.
\end{gather}
In Eqs. (\ref{Aright}) and (\ref{Aleft}), $\rho(\mathbf{r})$ denotes
the nuclear density (neutron plus proton 
contributions) and $A$ the mass of the fissioning nucleus. 
The neck abscissa $z_{\rm neck}$ is defined here as the $z$-value at which the 
nuclear density integrated in the perpendicular plane is 
minimal. This definition holds only for sufficiently necked-in 
nuclear shapes and corresponds to the value of $z$ for which the neck
radius is minimum. The light-fragment mass $A_L$ is thus simply given by 
$A_L=A-A_H$. The number of protons in the heavy and light 
fragments $Z_H$ and $Z_L$ are defined in the same way. 

Interpreting the fission modes in terms of valleys of the potential-%
energy surface, we do not need to explore the whole surface to
find these valleys and a constrained variational approach like the HFBCS 
approximation seems suitable to this purpose. However, two
limitations should be kept in mind. On the one hand, a partial
exploration of the energy surface might not lead to 
the lowest valleys at a given elongation (the driving coordinate here), on 
the other hand such an approach does not guarantee to find 
the lowest saddle point between a pair of local minima 
(in a given deformation space), as already pointed out in 
Ref.~\cite{papier_fission_PRC}. 
It will therefore be verified \textit{a posteriori}, 
by comparison with experiment, 
that the valleys obtained are the most relevant ones. An important point when 
using this approach is to take great care to check the stability of the 
solutions when looking for the relevant valleys. More precisely these 
solutions should correspond to significantly deep local minima 
for each constrained $Q_{20}$-value in the directions of mass asymmetry and 
neck coordinate. Finally, the point in a given fission valley 
where the solution corresponds to the larger constrained elongation 
is called the exit point of the fission valley. One can similarly 
define the exit point of a fusion valley upon considering the 
solution that corresponds to the smaller constrained elongation.

\section{Results}
\label{results}

The model and method presented in the previous section are applied to 
the $^{256}$Fm and $^{258}$Fm isotopes. Earlier calculations%
~\cite{papier_fission_EPJA} were performed for these heavy nuclei in the 
same framework except that left-right symmetry was imposed (in addition 
to axial symmetry). The results reported here are obtained by 
releasing this constraint. 

\begin{figure}[h]
\begin{center}
\includegraphics[width=8cm]{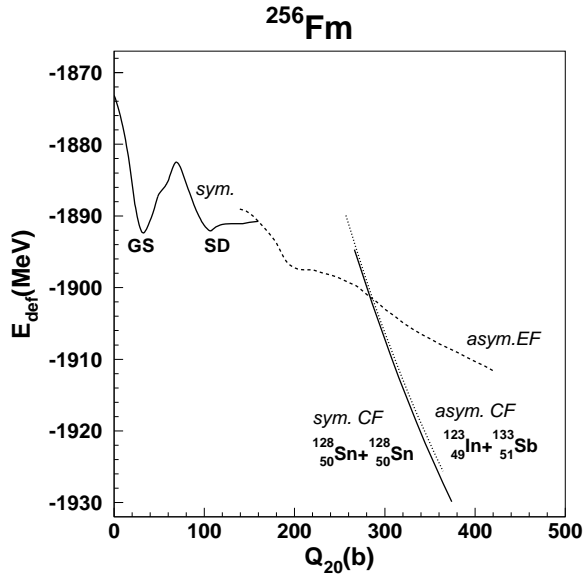}
\end{center}
\caption{Potential energy of deformation $E_{\rm def}$ as a function 
of the quadrupole moment $Q_{20}$ along the different fission
paths obtained for the $^{256}$Fm isotope. The solid line 
from 0 to 160 barns (b) represent left-right symmetric solutions
including the ground-%
state (GS) and the superdeformed (SD) minima. The solid and the dotted lines 
beyond 250~b correspond to symmetric and asymmetric compact fission (CF) 
paths, respectively. The dashed line is the bottom of the asymmetric 
elongated fission (EF) valley.
\label{fission_paths_Fm256}}
\end{figure}

For the $^{256}$Fm isotope, only left-right reflection symmetric 
solutions are found between the spherical shape at 
$Q_{20}=0$ and $Q_{20}\approx160$~b. The corresponding deformation-energy 
curve is plotted in Fig.~\ref{fission_paths_Fm256} as 
a solid line. A superdeformed minimum lies just beyond the inner fission 
barrier, around which preliminary calculations seem to 
indicate some softness in the direction of left-right asymmetric 
deformations (with a possible local minimum at a finite $Q_{30}$-value). 
Beyond this minimum, two kinds of valleys exist: the so-called 
fission valley, corresponding to one-body--shaped 
configurations, and the so-called fusion valley, corresponding to two 
separated fragments (see, for example, Refs.~\cite{Berger_QN,Berger_1984}). 

On the one hand, only one fission valley is obtained (dashed 
line labeled ``asym. EF'' -- asymmetric elongated fission -- in 
Fig.~\ref{fission_paths_Fm256}), along which 
the $^{256}$Fm nucleus exhibits left-right asymmetric shapes. The valley 
stretches from $Q_{20}\approx140$~b to $Q_{20}\approx420$~b, where the 
one-body--shaped solution becomes unstable against neck rupture. 
\begin{figure}[h]
\begin{center}
\includegraphics[width=8cm]{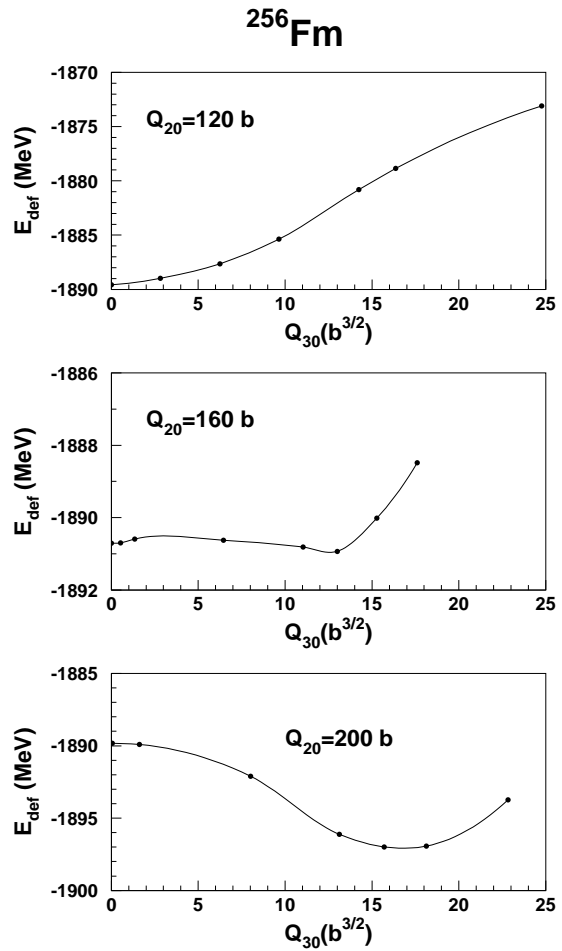}
\end{center}
\caption{Cuts in the potential-energy surface along the axial octupole 
moment $Q_{30}$ direction at 
different fixed elongations $Q_{20}$ for $^{256}$Fm. The full circles 
represent calculated points, whereas the solid lines are drawn as 
eye guides.\label{cutQ3_Fm256}}
\end{figure}
As we can see in Fig.~\ref{cutQ3_Fm256} showing slices of the potential-%
energy surface in the $Q_{30}$ direction at three different fixed elongations 
$Q_{20}$, the transition between the symmetric path from the superdeformed 
minimum and the asymmetric fission valley occurs smoothly around 
$Q_{20}=160$~b. At the exit point of the fission valley, the nascent 
fragments are calculated to be approximately $^{140}_{\;\:54}$Xe and 
$^{116}_{\;\:46}$Pd.

\begin{figure}[h]
\begin{center}
\includegraphics[width=8cm]{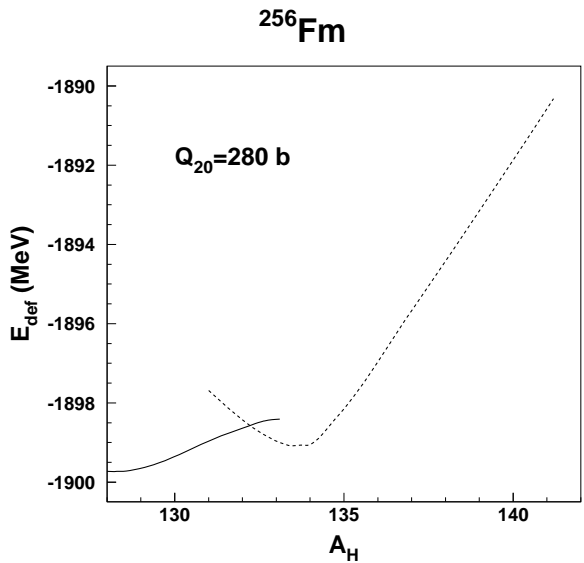}
\end{center}
\caption{Cut of the potential-energy surface along the $A_H$ direction at a 
fixed elongation $Q_{20}=280$~b for $^{256}$Fm. The solid line represents 
solutions obtained by increasing $A_H$ from 128 (symmetric solution) and the 
dashed line is obtained by decreasing $A_H$ from about 141.\label{cutAh_Fm256}}
\end{figure}

On the other hand, two fusion valleys corresponding to the symmetric 
$^{128}_{\;\:50}$Sn+$^{128}_{\;\:50}$Sn and slightly asymmetric 
$^{123}_{\;\:49}$In+$^{133}_{\;\:51}$Sb fragmentations are found in 
the potential-energy surface. They are represented in 
Fig.~\ref{fission_paths_Fm256} 
as a solid line and a dotted line, labeled ``sym. CF'' (symmetric
compact fission) and ``asym. CF'' (asymmetric compact fission), 
respectively. Although these valleys are plotted only for elongations
less than $Q_{20}=375$~b, they exist for larger elongations. In
contrast their upper ends represent exit points 
as defined in the end of the previous section. The two fusion valleys can 
also be visualized in Fig.~\ref{cutAh_Fm256} showing
a cut in the heavy-fragment mass $A_H$ direction at a fixed elongation 
$Q_{20}=280$~b chosen as an example. 
In this figure, the solid line represents solutions obtained by 
increasing the constraint on $A_H$ from 128 (symmetric solution) to 
about 133, beyond which these solutions become unstable. Similarly the 
dashed line corresponds to solutions obtained by decreasing the constraint 
on $A_H$ from about 141 to about 131, below which 
this kind of solutions does not exist. The two minima at 
$A_H=128$ and $A_H=133.5$ are therefore not connected through 
a continuous curve in Fig.~\ref{cutAh_Fm256}, from which we learn that one or 
several additional degrees of freedom are missing in the description of 
this region of the energy surface (for example the fragment deformations). 
Nevertheless this does not cast any doubt on the existence of the two fusion 
valleys. Regarding the exact location of the asymmetric minimum on the 
dashed curve of Fig.~\ref{cutAh_Fm256}, the
present HFBCS calculations indicate that the mass of the heavy fragment varies 
very little in the asymmetric fusion valley, with an average integer value 
of 133.

\begin{figure}[h]
\begin{center}
\includegraphics[width=8cm]{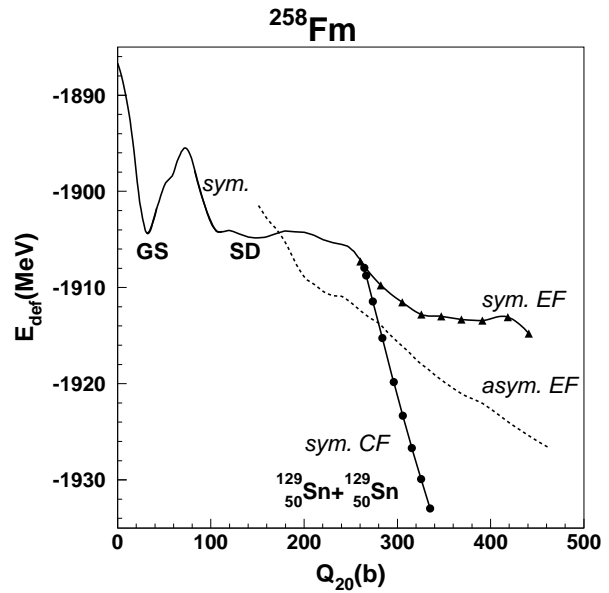}
\end{center}
\caption{Same as Fig.~\ref{fission_paths_Fm256} for the $^{258}$Fm isotope. 
Full circles and triangles are calculated points along the symmetric compact
fission (``sym. CF'') and symmetric elongated fission (``sym. EF'')
paths, respectively.
\label{fission_paths_Fm258}}
\end{figure}

Let us now turn to the $^{258}$Fm isotope. Its potential-energy surface 
presents some features in common with the one of $^{256}$Fm 
from the spherical point to the also present superdeformed minimum, 
with only symmetric solutions -- corresponding to the solid line in 
Fig.~\ref{fission_paths_Fm258} -- in this elongation range. 
As in $^{256}$Fm, octupole softness is observed around the top of the inner 
fission barrier. An asymmetric fission valley emerging in the vicinity 
of the superdeformed minimum is also present in $^{258}$Fm and plotted 
as a dashed line in Fig.~\ref{fission_paths_Fm258}. The nascent 
fragmentation around the exit point, namely approximately 
$^{141}_{\;\:54}$Xe+$^{117}_{\;\:46}$Pd, is similar to that 
obtained in $^{256}$Fm. However there are two major differences
between $^{256}$Fm and $^{258}$Fm. The symmetrical solutions in $^{258}$Fm not 
only remain stable against left-right asymmetric deformations until 
scission, but they also give rise to two different families 
of nuclear shapes. The solutions with compact shapes constitute the 
symmetric compact fission (CF) path marked with full circles in 
Fig.~\ref{fission_paths_Fm258} (labeled ``sym. CF'') and those 
associated with elongated configurations form the symmetric elongated 
fission (EF) path marked with full triangles (labeled ``sym. EF''). 

A word of caution should be said here about the 
symmetric EF path. As mentioned in Sect.~2, the present 
calculations are done using a harmonic-oscillator basis size $N_0=16$ 
for the expansion of single-particle states. Even though this is large 
enough to show the existence of the 
symmetric EF path (that is, its stability with respect to 
left-right asymmetric deformations), it is not the case when addressing 
quantitative questions like the location of the exit point. This is why axial 
and left-right symmetric calculations are performed with a larger 
basis size $N_0=20$. It then becomes possible to scan the potential-energy 
surface on a mesh $(\Dcm,Q_{\rm 2f})$, where $\Dcm$ and $Q_{\rm 2f}$ stand, 
respectively, for the center-of-mass distance and the fragment axial 
quadrupole moment (which is of course the same for each fragment because 
of the reflection symmetry). Since we deal here with very large elongations 
of the fissioning nucleus, the centers of mass of the pre-fragments (and 
\textit{a fortiori} of the separated fragments) are well defined and 
it becomes more physical and intuitive to consider $\Dcm$ rather 
than $Q_{20}$. It is worth adding here that $Q_{20}$, $Q_{\rm 2f}$ and 
$\Dcm$ are not independent of each other since they obey the relation
\eq{
Q_{20}=2\,Q_{\rm 2f}+\frac{A}{2}\,\Dcm^2\:,
}
where $A$ is the mass number of the fissioning nucleus. 

In practice, it is extremely time consuming to optimize the basis parameters 
$b$ and $q$ at each grid point. Therefore I~resort to an approximate
procedure in which $b$ is optimized using a smaller basis
corresponding to $N_0=16$ at an elongation $Q_{20}\approx190$~b
somewhat smaller than the elongation at which the symmetric EF valley 
appears. The resulting value of $b$ is then used over the whole range of 
deformations covered by the mesh ($10\mbox{ fm}\leqslant \Dcm 
\leqslant 23\mbox{ fm}$ and $-30\mbox{ b}\leqslant Q_{\rm 2f} 
\leqslant 60\mbox{ b}$). As we can see in Tab.~\ref{opt190b_Fm258}, 
0.42 is the approximate optimal value for $b$.
\begin{table}[h]
\caption{Values of the deformation energy on a $(b,q)$ mesh ($b$ in 
$\rm fm^{-1}$) at $Q_{20}=190$~b for $^{258}$Fm, calculated with 17 major 
shells ($N_0=16$).\label{opt190b_Fm258}}
\begin{center}
\begin{tabular}{ccccccccccc}
\hline
\hline
& \backslashbox{$b$}{$q$} && 1.6 && 1.8 && 2.0 && 2.2 & \\
\hline
& 0.38 && $-1903.03$ && $-1903.28$ && $-1903.75$ && $-1903.67$ & \\
& 0.40 && $-1903.85$ && $-1904.04$ && $-1904.21$ && $-1903.93$ & \\
& 0.42 && $-1903.99$ && $-1904.05$ && $-1904.21$ && $-1903.66$ & \\
& 0.44 && $-1903.61$ && $-1903.66$ && $-1903.91$ && $-1903.37$ & \\
& 0.46 && $-1903.15$ && $-1903.31$ && $-1903.56$ && $-1903.16$ & \\
\hline
\hline
\end{tabular}
\end{center}
\end{table}
As for $q$, to which the results are not very sensitive 
(see Tab.~\ref{opt190b_Fm258}), especially with the enlarged basis, an 
approximate variation with $\Dcm$ is taken into account (the actual 
values of $q$ appear in Fig.~\ref{cutPESFm258} discussed 
below). The resulting numerical uncertainties are not expected 
to drastically affect the relative position of 
the symmetric EF and CF valleys or the nuclear shape at the exit 
point of the EF valley.

\begin{figure}[h]
\hspace*{-0.65cm}
\includegraphics[scale=0.75]{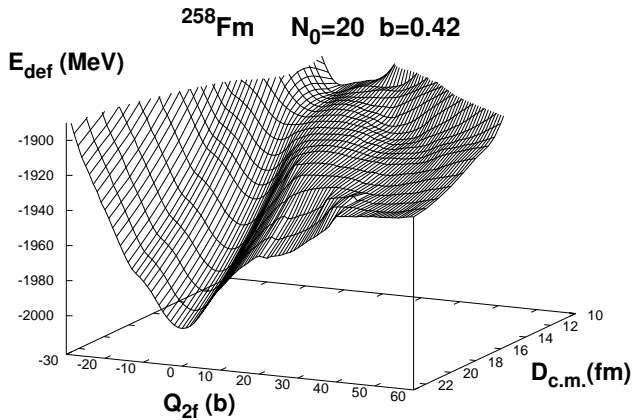}
\caption{Surface of deformation energy as a function of the center-of-mass 
distance $\Dcm$ and the fragment elongation $Q_{\rm 2f}$ for 
$^{258}$Fm.\label{PES_Dcm_Q2f}}
\end{figure}

\begin{figure}[h]
\vspace*{-1cm}
\hspace*{-2cm}
\includegraphics[scale=0.85]{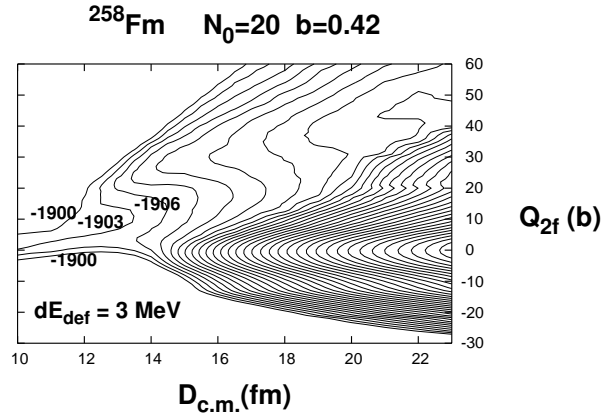}
\vspace*{-1cm}
\caption{Contour diagram of the deformation energy as a function of the 
center-of-mass distance $\Dcm$ and the fragment elongation 
$Q_{\rm 2f}$ for $^{258}$Fm. The energy interval between the contour lines 
($\rm dE_{def}$) is 3 MeV.\label{contoursEtot_Dcm_Q2f}}
\end{figure}

\begin{figure*}[h]
\begin{center}
\includegraphics[width=17cm]{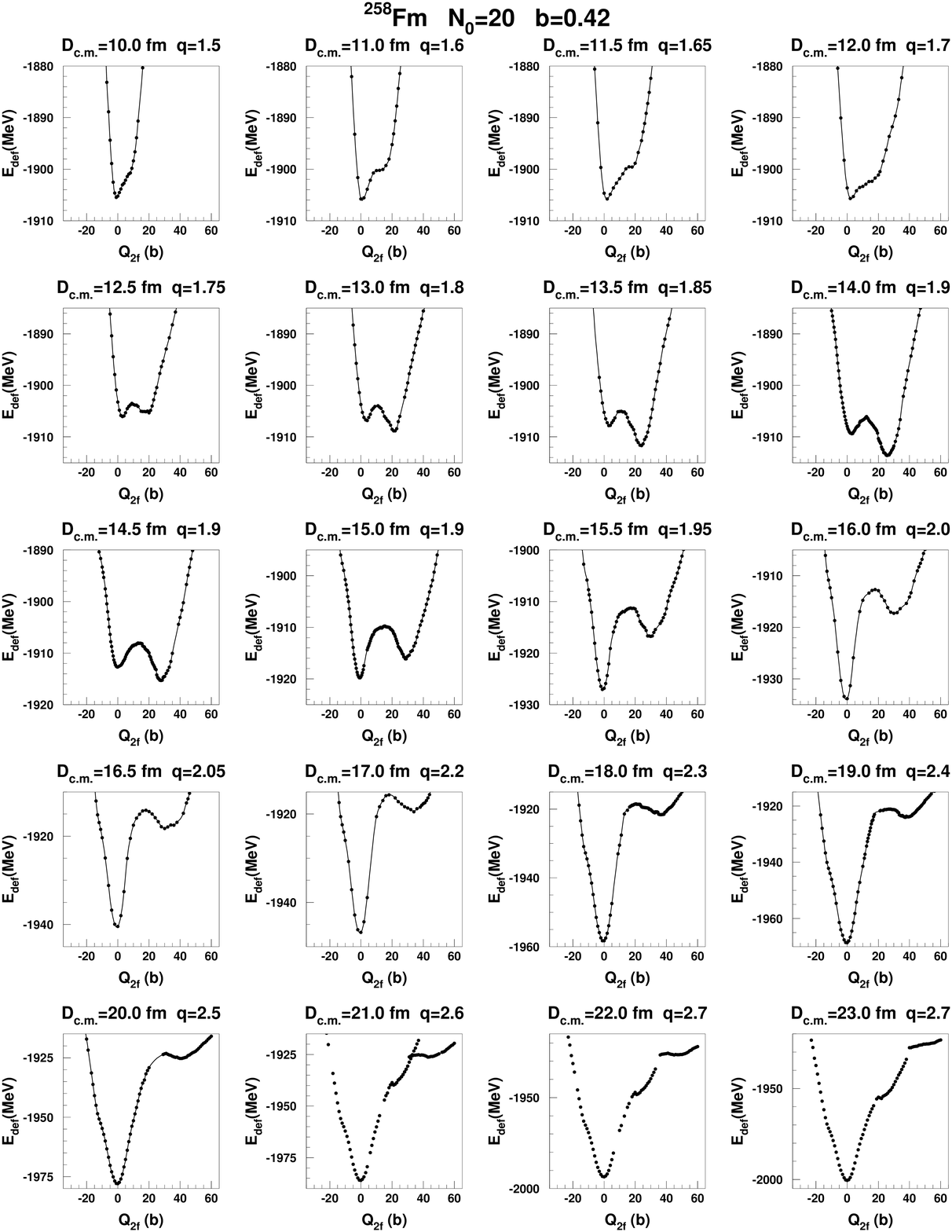}
\end{center}
\caption{Cuts in the potential-energy surface along the $Q_{\rm 2f}$ direction at 
different fixed center-of-mass distances $\Dcm$ for $^{258}$Fm.
\label{cutPESFm258}}
\end{figure*}

The potential-energy surface obtained for $^{258}$Fm is displayed in 
three different forms: as a three-dimensional surface in 
Fig.~\ref{PES_Dcm_Q2f}, as a two-dimensional contour diagram in 
Fig.~\ref{contoursEtot_Dcm_Q2f} and as a series of cuts at 
fixed $\Dcm$-values in Fig.~\ref{cutPESFm258}.
The deformation-energy curves as functions of $Q_{\rm 2f}$ in 
Fig.~\ref{cutPESFm258} exhibit two minima for $\Dcm\geqslant 12.5$~fm. 
One at $Q_{\rm 2f}\approx 0$ is associated with a configuration having very 
few nucleons in the neck ($Q_N\approx 0$), corresponding 
thus to two separated, identical and nearly spherical fragments (symmetric 
CF valley). The other minimum varies with $\Dcm$ from 18~b to 45~b and 
the associated configuration has a finite neck radius ($Q_N\geqslant 6$): 
it corresponds to the symmetric EF valley. The deformation-energy
curves are all continuous except for 
$\Dcm\geqslant 21$~fm (for which only the calculated points are
plotted). Indeed two kinds of configurations coexist 
for $30\mbox{ b}\leqslant Q_{\rm 2f}\leqslant 40\mbox{ b}$. They are 
characterized by different $Q_N$-values: $Q_N\approx 0$ for the steep 
increasing branch (separated fragments) and $Q_N\approx 6$ 
for the other one (very elongated one-body-shaped configuration). From 
Fig.~\ref{cutPESFm258} we can approximately localize the exit point of 
the symmetric EF valley at about $\Dcm=22$~fm, with $Q_{\rm 2f}\approx 45$~b.

\begin{figure}[h]
\begin{center}
\includegraphics[height=10cm]{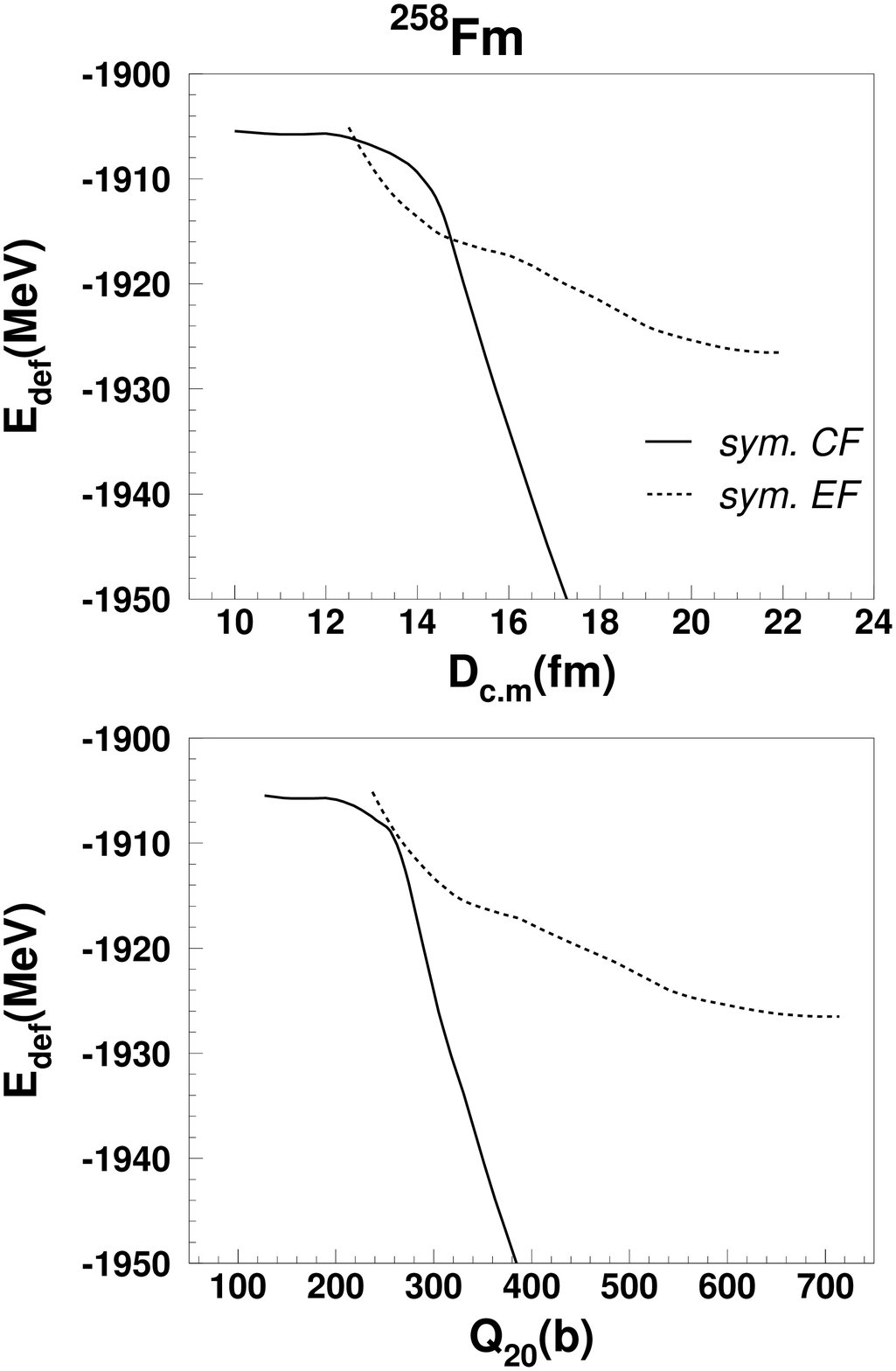}
\end{center}
\caption{Deformation energy along both symmetric fission valleys as a 
function of $\Dcm$ (upper panel) and as a function of $Q_{20}$ 
(lower panel), obtained with $N_0=20$ and $b=0.42$.
\label{sym_valleysFm258_Edef}}
\end{figure}

\begin{figure}[h]
\begin{center}
\includegraphics[height=8cm]{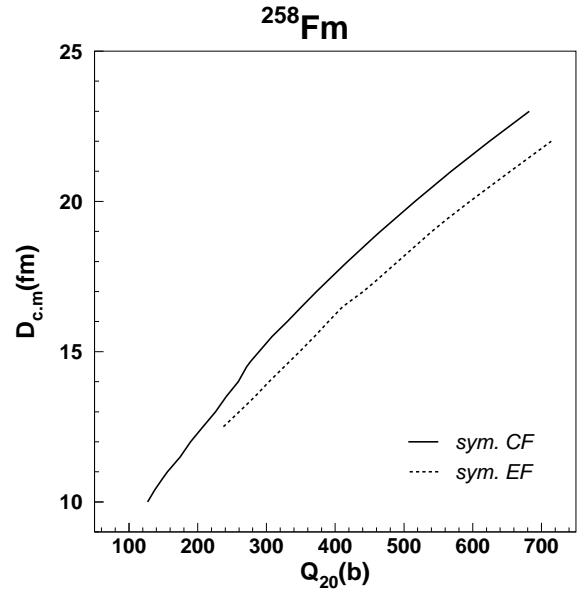}
\end{center}
\caption{Variation of $\Dcm$ along both symmetric fission valleys as a 
function of $Q_{20}$, obtained  with $N_0=20$ and $b=0.42$.
\label{sym_valleysFm258_Dcm_Q20}}
\end{figure}

From these results, we can deduce the variation of the deformation energy 
along the bottom of each valley as a function of the driving
coordinate $\Dcm$ (see the upper panel of Fig.~\ref{sym_valleysFm258_Edef}). 
It is also instructive to plot the same curves as functions of the 
quadrupole moment $Q_{20}$ (see the lower panel of 
Fig.~\ref{sym_valleysFm258_Edef}). 
The variation of the energy is of course not affected by the choice of 
the driving coordinate since $\Dcm$ is a monotonically 
increasing function of $Q_{20}$ as shown in 
Fig.~\ref{sym_valleysFm258_Dcm_Q20}, but the relative position of the two 
valleys can differ because of a projection effect. Indeed projecting 
the multidimensional energy surface onto a deformation subspace distorts 
it and the resulting pattern generally depends on the actual subspace. 
However, we can neglect this effect when comparing two valleys if the 
distortion is weak or the energy difference between the two valleys 
is large.

\section{Discussion}

\subsection{Asymmetric-to-symmetric transition in the fragment-mass 
distribution}

From the features of the potential-energy surface we can obtain some 
information about the fragment-mass distribution in the spontaneous fission 
of $^{256}$Fm and $^{258}$Fm. 

Starting from the ground state of $^{256}$Fm, the lowest and only 
continuous path leading to scission is asymmetric in its late stages, 
where the nascent fragments have a fairly constant mass ratio of 
$A_H/A_L\approx 140/116$ beyond $Q_{20}\approx 350$~b. 
Under the assumption that the most probable fragmentation experimentally 
observed corresponds to the configuration just before neck
rupture, that is, at the exit point of the fission valley, 
the mass distribution in the spontaneous fission of 
$^{256}$Fm is inferred to be asymmetric and peaked at 
$A_L\approx116$ and $A_H\approx140$, only one mass unit away from the
experimental value for the heavy fragment $A_H\approx141$
\cite{Flynn}. This property 
has also been successfully described within the macroscopic-microscopic 
FRLDM model by M{\"o}ller \textit{et al.}~\cite{Moller_Nature} who 
found $A_H\approx140$. Moreover, 
the calculations by Warda \textit{et al.}~\cite{Warda2002} have shown the 
same behavior in the transition from the ground-state symmetric path to the 
asymmetric fission valley. This can be seen in their 
Fig.~5 where their cuts in the $Q_N$ direction at various elongations 
$Q_{2}=Q_{20}/2$ are similar to the HFBCS cuts along
$Q_{30}$, since $Q_N$ and $Q_{30}$ turn out to be in a one-to-one 
correspondence along the cuts of Fig.~\ref{cutQ3_Fm256} here. It is 
worth mentioning that the portion of the symmetric path between 
$Q_2\approx90$ and $Q_2\approx130$~b of 
their Fig.~3 does not correspond to local minima in the $Q_N$ 
direction, as can be seen in their Fig.~5. Although the authors of
Ref.~\cite{Warda2002} found a clear left-right reflection asymmetry 
in their nuclear shapes along the EF path, they reported that the two 
nascent fragments have nearly equal masses, which they inferred in an 
unclear way from the integrated particle number as a function 
of $z$ plotted in their Fig.~4b. 
On the contrary, assuming that the neck most likely ruptures where its radius 
is minimum, the Figs.~4a and 4b of Ref.~\cite{Warda2002} 
rather indicate that the heavy fragment has a mass of about 136 and an 
atomic number of about 52, in a much better agreement with the
HFBCS results and the experimental data~\cite{Flynn}.

Contrary to $^{256}$Fm, the most favorable exit channel (i.e, the
lowest continuous path) beyond the superdeformed minimum in $^{258}$Fm
goes along the only symmetric path which eventually forks, instead of 
following the asymmetric EF valley because of the ridge separating
them (at least 1.5~MeV high according to Warda \textit{et
al.}~\cite{Warda2002}). Whichever valley is 
eventually followed by the spontaneously fissioning nucleus towards 
scission, the outcome is the same in terms of mass fragmentation 
since in either valley the configurations are left-right symmetric. 
This leads thus to a symmetric mass distribution, as was obtained 
in all the other theoretical sudies and in agreement with 
experiment~\cite{Hoffman_systematics,Hulet_Fm258_PRL1986}. 

\subsection{Bimodal fission in $^{258}$Fm}

Let us now relate the properties of the valleys obtained in the 
potential-energy surface of $^{258}$Fm to the fragment total-kinetic-energy
and mass distributions of this isotope. 

\begin{figure}[h]
\begin{center}
$\Dcm=13.0$~fm, $Q_{20}=226.02$~b \\
\vspace*{-0.5cm}
\includegraphics[width=6.5cm]{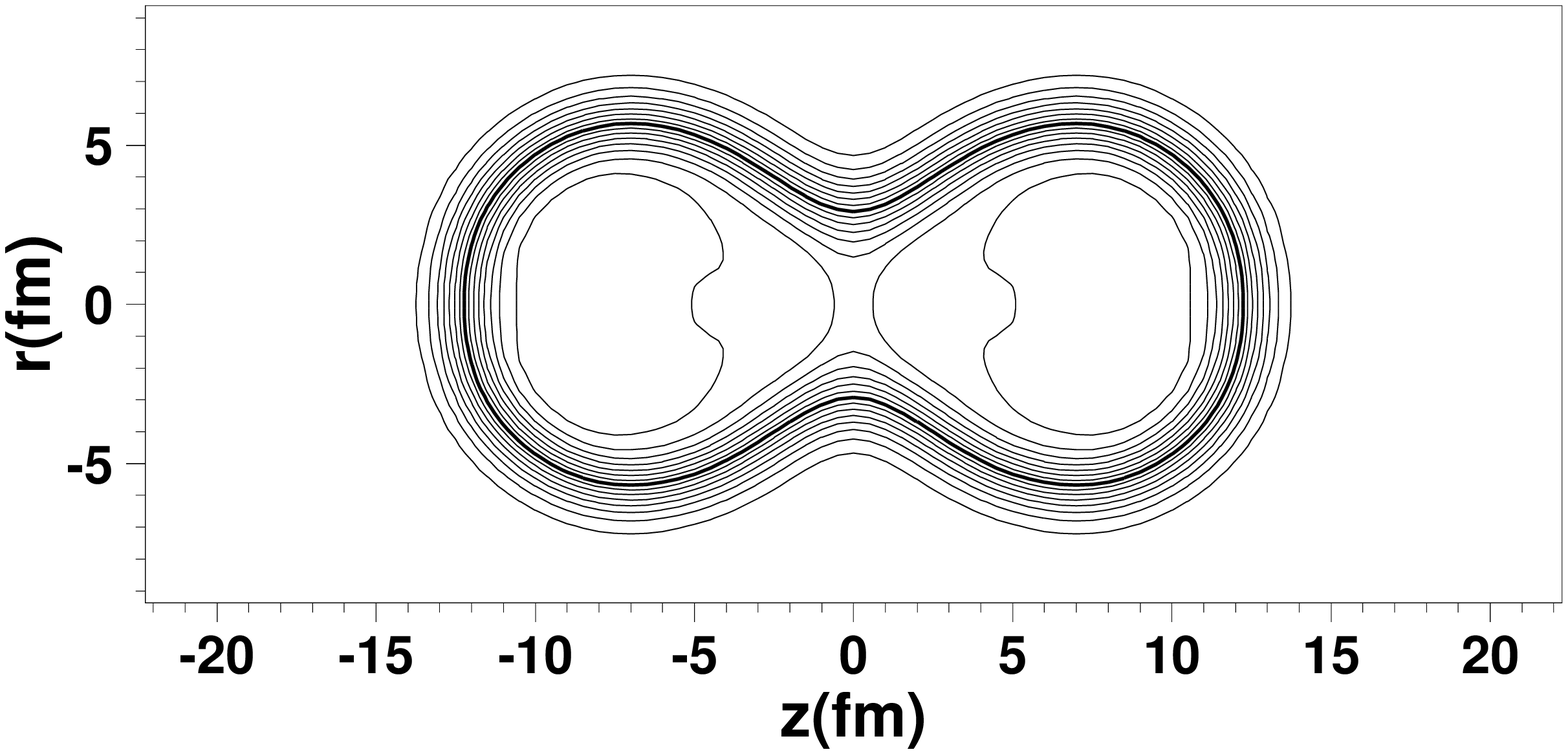}\\
$\Dcm=14.0$~fm, $Q_{20}=254.85$~b \\
\vspace*{-0.5cm}
\includegraphics[width=6.5cm]{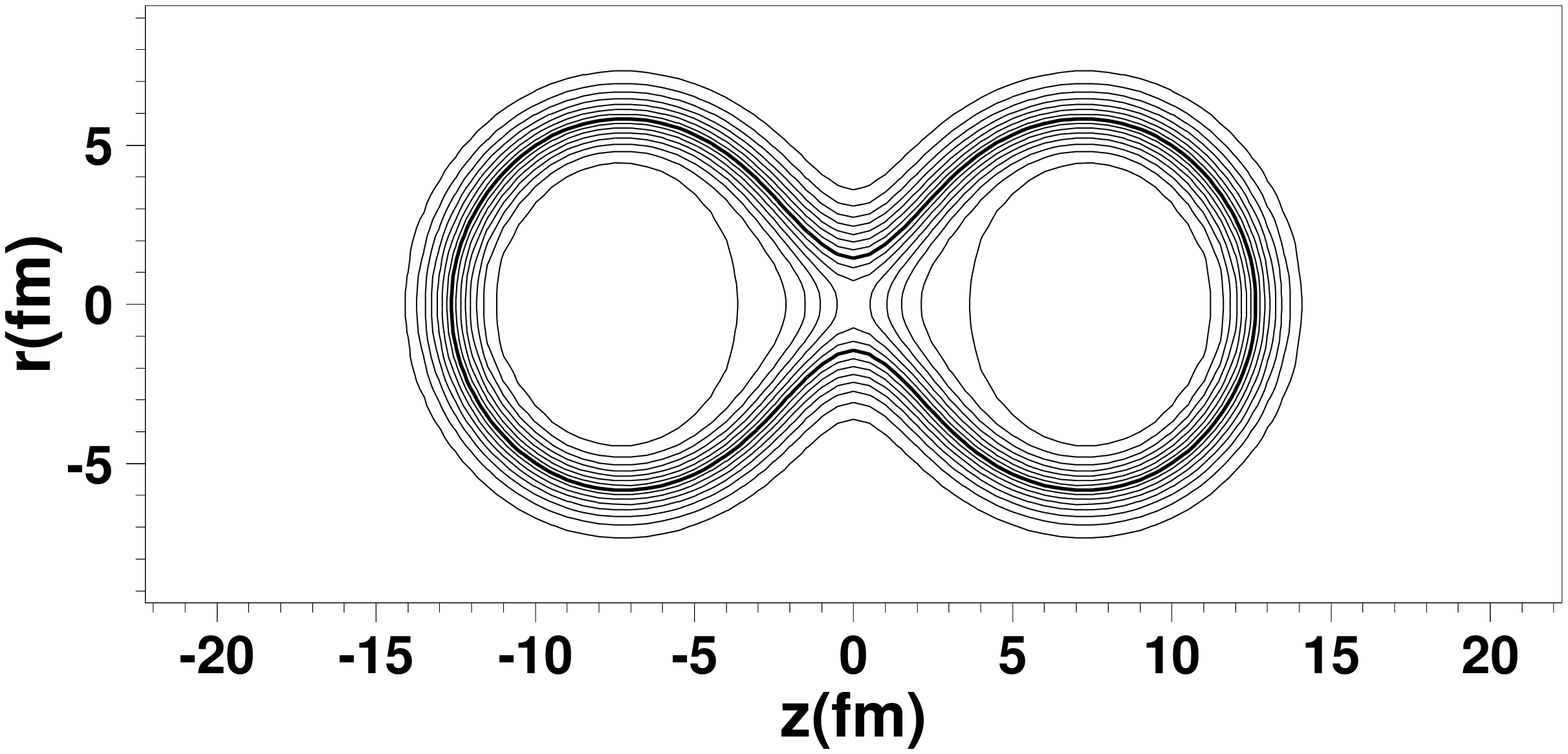}\\
$\Dcm=14.5$~fm, $Q_{20}=271.23$~b \\
\vspace*{-0.5cm}
\includegraphics[width=6.5cm]{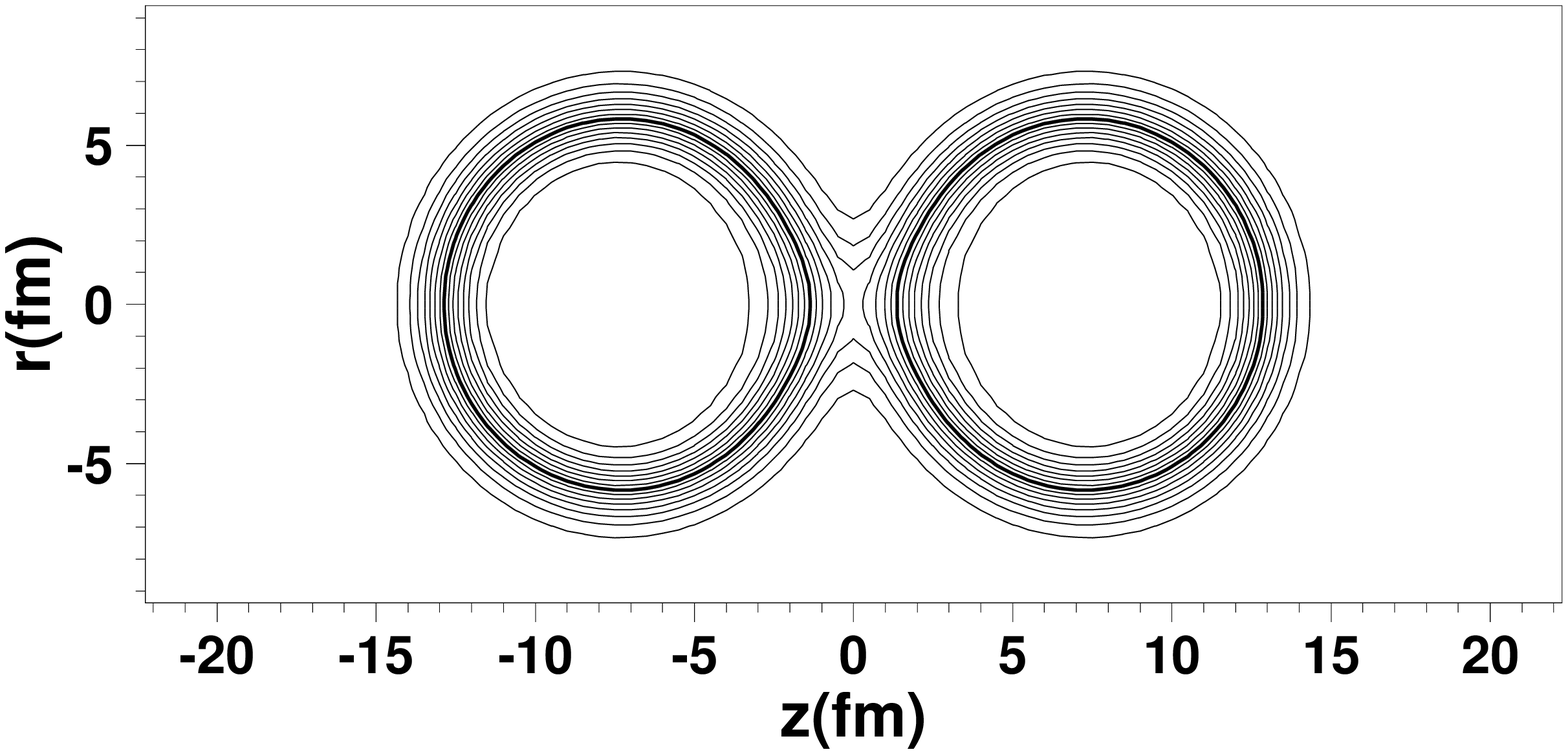}\\
$\Dcm=15.0$~fm, $Q_{20}=289.26$~b \\
\vspace*{-0.5cm}
\includegraphics[width=6.5cm]{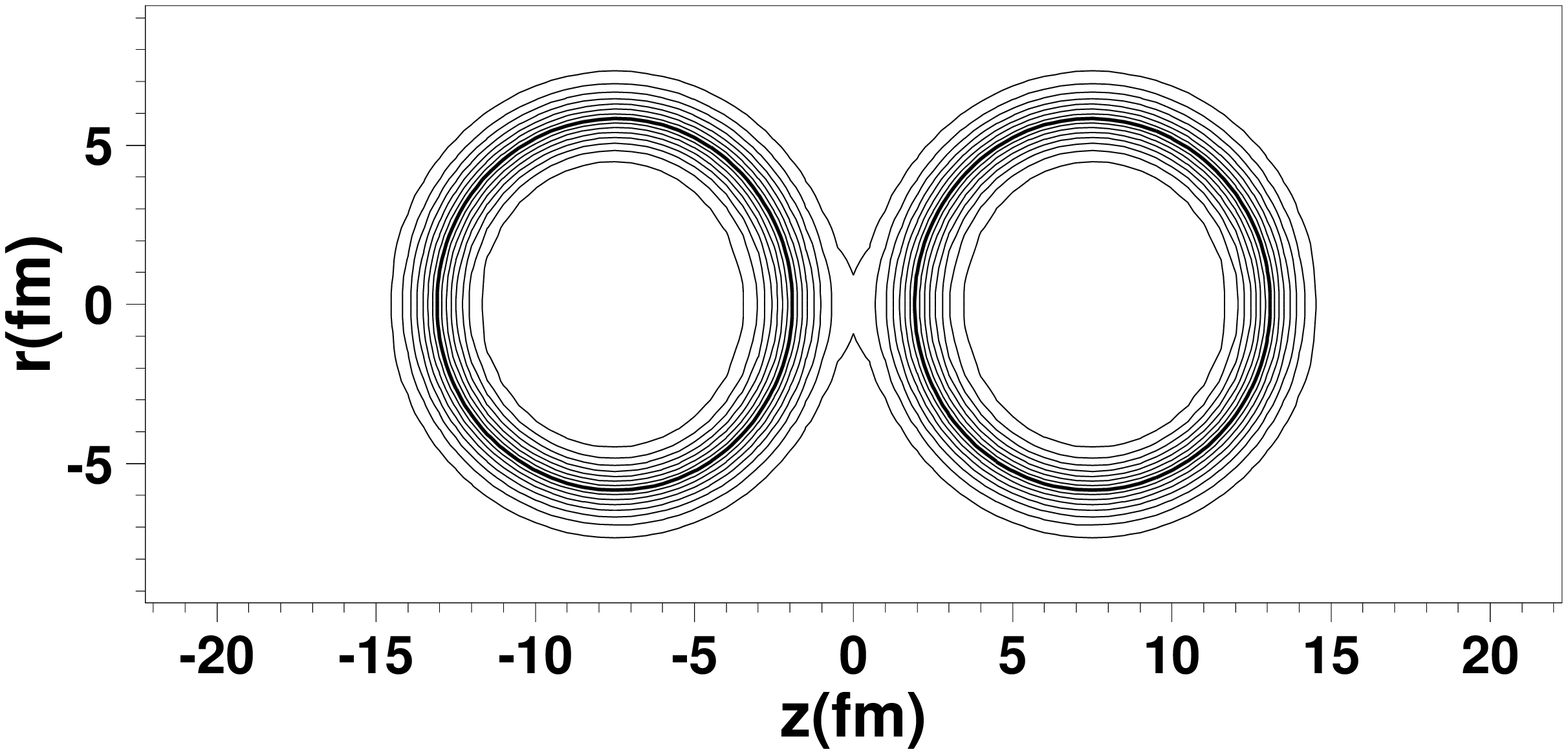}
\end{center}
\caption{Sequence of nuclear shapes along the symmetric CF valley of 
$^{258}$Fm between $\Dcm=13.0$~fm and $\Dcm=15.0$~fm. The solid 
lines correspond to equal nuclear density contours ranging from 
$\rho(\mathbf{r})=0.01 \mbox{ fm}^{-3}$ (outermost contour) to 
$\rho(\mathbf{r})=0.15 \mbox{ fm}^{-3}$ (innermost contour) with 
$0.01 \mbox{ fm}^{-3}$ steps. The thick solid 
line marks the density contour at half the saturation density, namely 
$\rho(\mathbf{r})=\rho_{\rm sat}/2=0.08 \mbox{ fm}^{-3}$.
\label{shapesFm258CFsym}}
\end{figure}

\begin{figure}[h]
\begin{center}
$\Dcm=12.5$~fm, $Q_{20}=237.57$~b \\
\vspace*{-0.5cm}
\includegraphics[width=6.5cm]{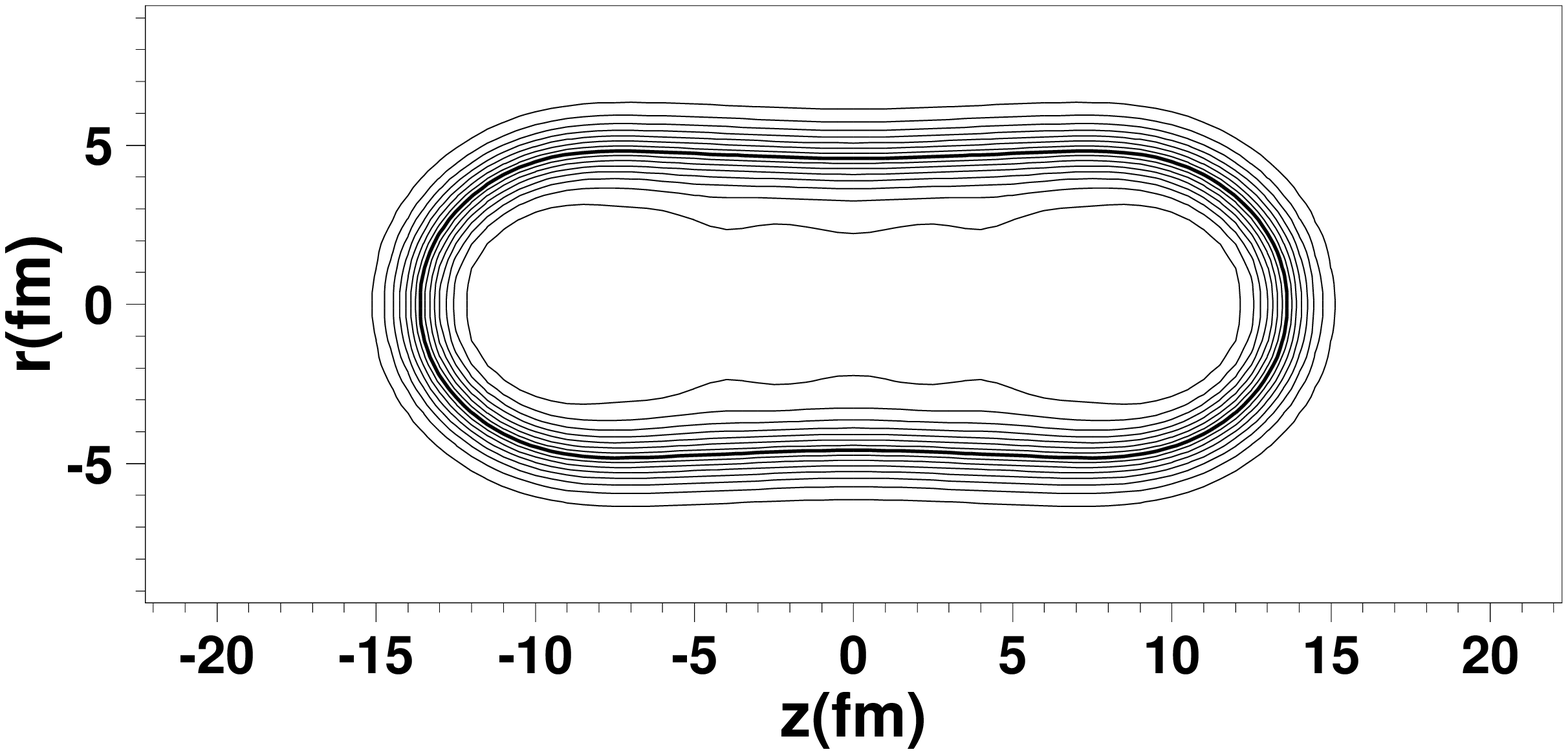}\\
$\Dcm=17.0$~fm, $Q_{20}=440.82$~b \\
\vspace*{-0.5cm}
\includegraphics[width=6.5cm]{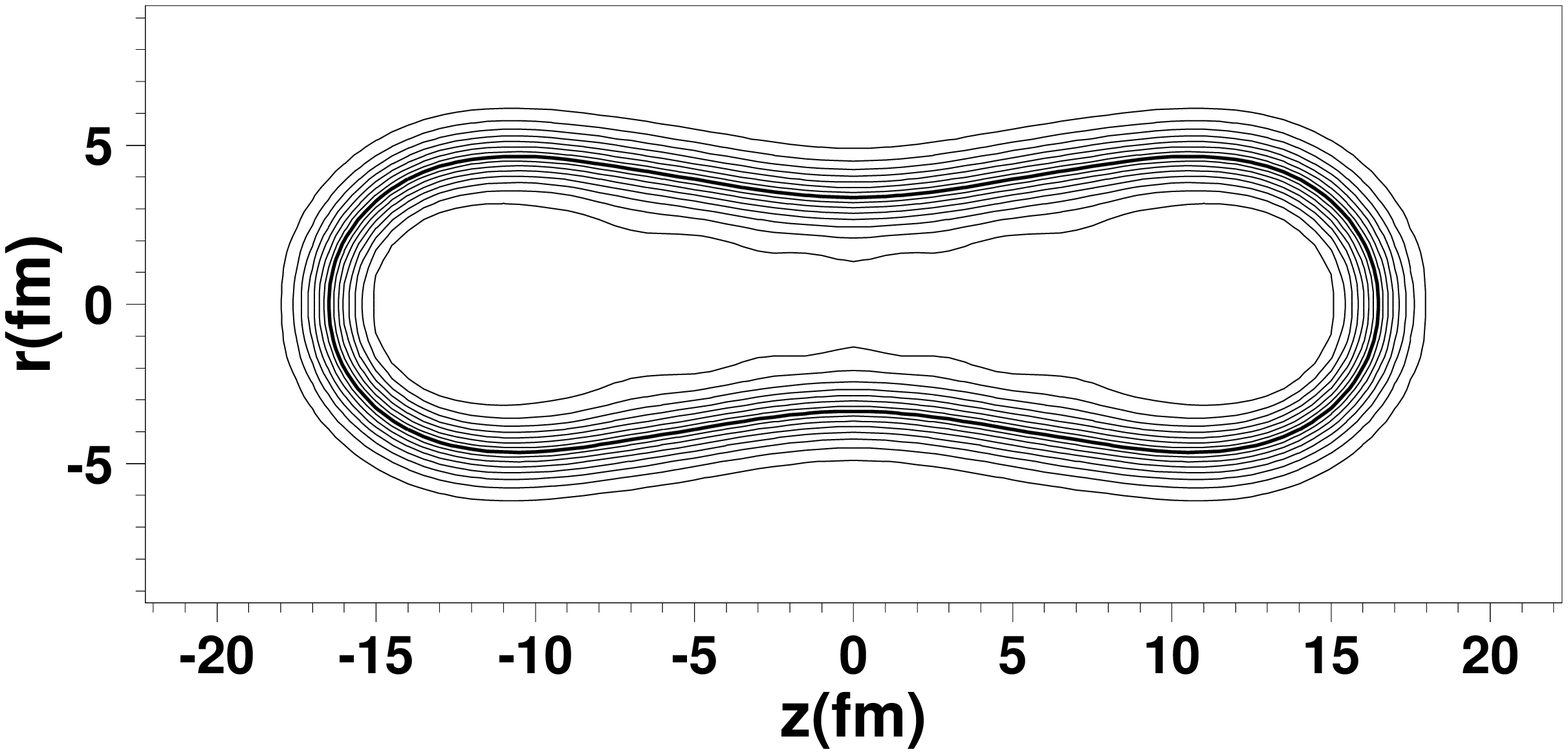}\\
$\Dcm=20.0$~fm, $Q_{20}=596.02$~b \\
\vspace*{-0.5cm}
\includegraphics[width=6.5cm]{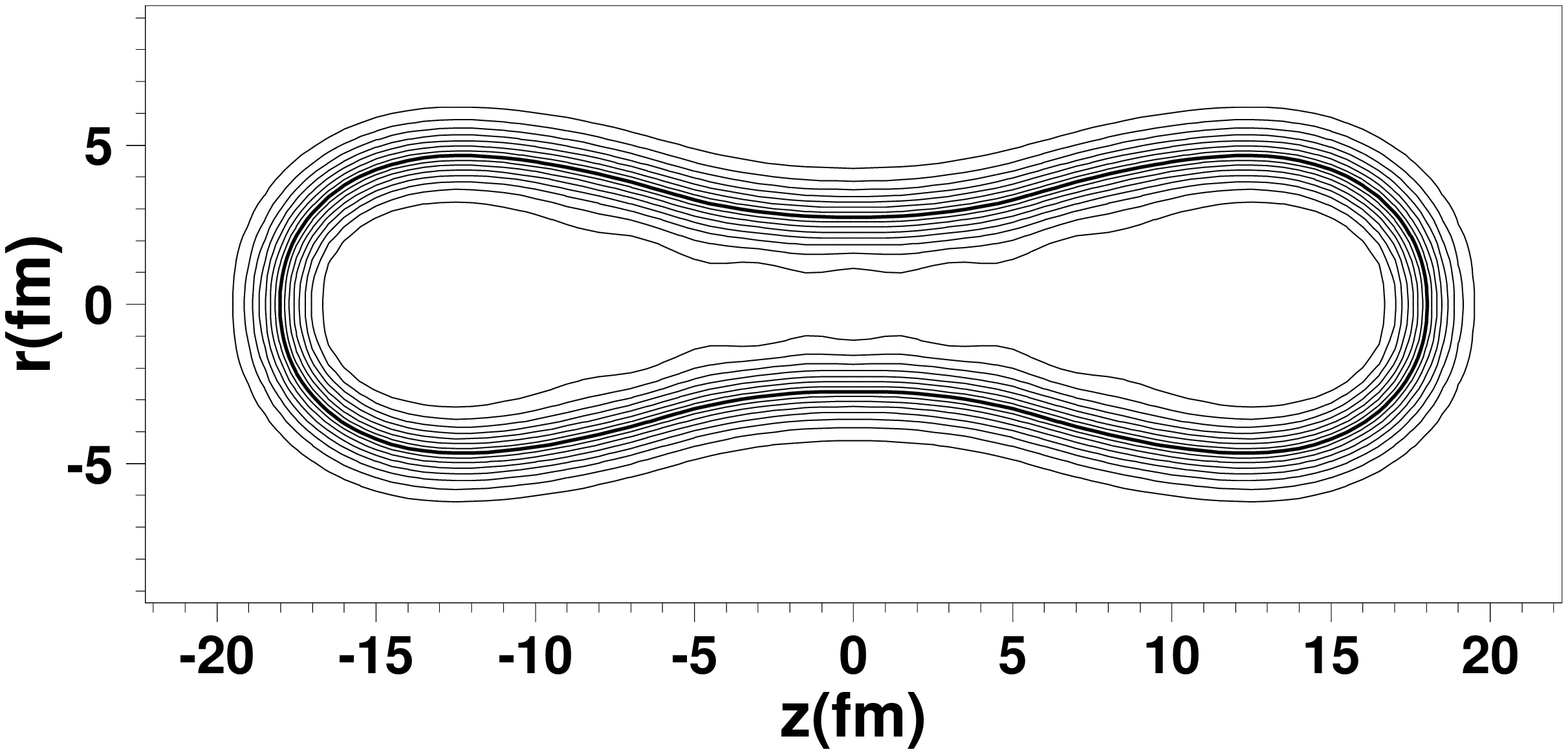}\\
$\Dcm=22.0$~fm, $Q_{20}=714.36$~b \\
\vspace*{-0.5cm}
\includegraphics[width=6.5cm]{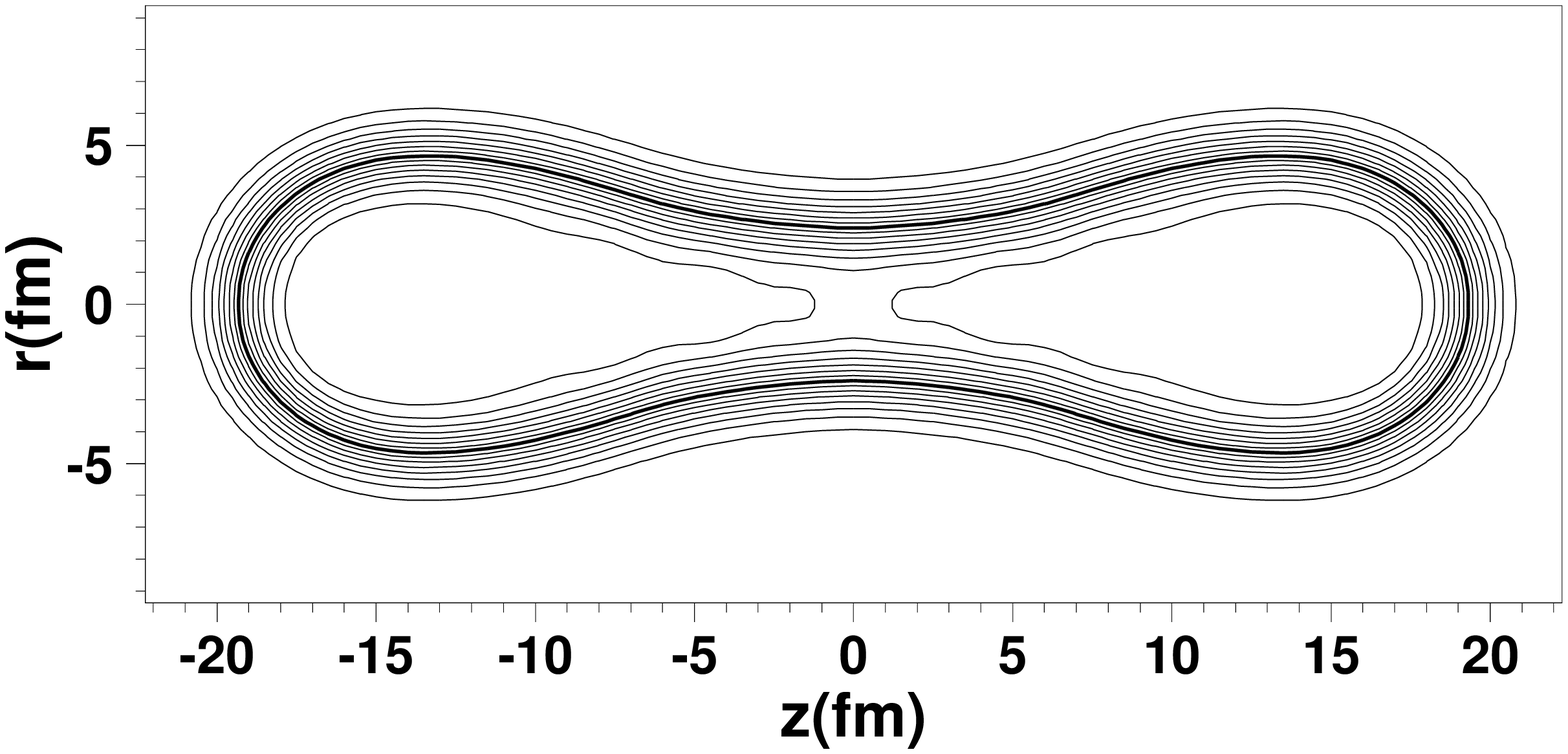}\\
\end{center}
\caption{Same as Fig.~\ref{shapesFm258CFsym} for nuclear shapes along the 
symmetric EF valley between $\Dcm=12.5$~fm and $\Dcm=22.0$~fm.
\label{shapesFm258EFsym}}
\end{figure}

The two symmetric valleys present a major difference associated with 
the nuclear shapes. Whereas the fissioning nucleus develops already 
in the early stages of the CF valley a narrow neck connecting two nearly 
spherical nascent fragments (see Fig.~\ref{shapesFm258CFsym}), 
the neck of a fissioning nucleus following the symmetric EF valley
persists over a much wider range of total elongation $Q_{20}$, 
with very elongated nascent fragments (see Fig.~\ref{shapesFm258EFsym}).
From a geometrical argument and approximating the total fragment 
kinetic energy by the Coulomb interaction 
energy (the dominant contribution) at scission, we can deduce that the 
fragments formed in the descent of the fissioning nucleus along the 
symmetric CF valley have a much higher kinetic energy than those 
associated with fission events from the symmetric EF valley. 
In both cases the fragment-mass distribution is expected to be
symmetric since the parent nucleus fissions into two identical fragments
in both valleys. More specifically, in the case where $^{258}$Fm 
undergoes fission through compact shapes like the ones displayed in 
Fig.~\ref{shapesFm258CFsym}, the high stiffness of the CF valley in
the mass asymmetry direction, resulting from the strong shell effects 
in the nearly spherical and doubly magic nascent fragments, produces a 
mass distribution much narrower than the one corresponding to
fission events through the symmetric EF valley.

These arguments show that the two HFBCS symmetric valleys are
consistent with the kinetic-energy and mass-distribution properties 
of the two modes experimentally observed. 
Therefore it is natural to identify the symmetric EF path with the 
low-TKE mode, and the symmetric CF path with the high-TKE mode. Another 
argument in favor of the interpretation of the symmetric EF valley relies 
on the similarity between the nuclear shapes of Fig.~\ref{shapesFm258EFsym} 
and those obtained in the liquid-drop model, with which the experimentally 
measured properties of the low-energy mode are 
consistent~\cite{Hulet_Fm258_PRC1989}. 

To provide more quantitative grounds to this interpretation, 
a discussion is now devoted to the estimation of the fragment total 
kinetic energy associated with each fission path. Let us assume that TKE 
is given by
\eq{
{\rm TKE}=E_{\rm C}^{(\rm int)}+E_{\rm N}^{(\rm int)}+E_{\rm K}^{(\rm sc)}\:,
}
where a small nuclear interaction energy 
$E_{\rm N}^{(\rm int)}$ and a somewhat larger prescission kinetic energy 
$E_{\rm K}^{(\rm sc)}$ are added to the dominant Coulomb 
interaction energy $E_{\rm C}^{(\rm int)}$. These three contributions 
have to be calculated just after neck rupture, that is, at
scission. The Coulomb part does not pose any particular problem, so I
focus on how to calculate the prescission kinetic
and nuclear contributions. The relative energy of a scission 
configuration $\Delta E_{\rm sc}$ with respect to the initial energy of the 
fissioning nucleus (its ground-state energy in the present case of 
spontaneous fission) represents the available energy at scission. 
This energy can be shared among collective degrees of freedom 
other than deformation, essentially kinetic energy in the fission 
direction $E_{\rm K}^{(\rm sc)}$, and internal degrees of
freedom as an ``internal'' excitation energy $E_{\rm T}^{(\rm sc)}$. 
We can thus write:
\eq{
-\Delta E_{\rm sc}=E_{\rm K}^{(\rm sc)}+E_{\rm T}^{(\rm sc)}\:,
}
assuming that a physical scission configuration 
lies below the ground state to be accessible by tunneling from the 
ground-state well ($\Delta E_{\rm sc}<0$). Since a static study is unable 
to provide the actual partitioning, I postulate an equipartition of 
$-\Delta E_{\rm sc}/2$ between its two components
\eq{
\label{Ekinpresc}
E_{\rm K}^{(\rm sc)}=E_{\rm T}^{(\rm sc)}\approx-\frac{\Delta E_{\rm sc}}{2}\:.
}
This approximation leads to values of prescission kinetic energy of 
about 10~MeV for the scission configurations considered in 
Tab.~\ref{scission_Fm258}. This is of the same 
order as the one assumed by Brosa \textit{et al.}~\cite{Brosa_PhysRep1990} 
and consistent with the experimental average TKE-values, as well as
the one found by Bonasera \cite{Bonasera} in semi-classical 
dynamical calculations and by Abe \textit{et al.}~\cite{Abe} from dynamical 
calculations based on the two-dimensional Langevin equation and the one-body 
dissipation mechanism. As for the nuclear interaction between 
the two fragments, it can be in principle calculated with the 
Skyrme force. This requires one to disentangle the three contributions
of the nuclear energy of the whole system, namely the self-energies of
the two fragments and the interaction energy between the fragments. This is 
ambiguous since one has to unfold 
the local densities into two sets of densities localized each in one 
of the fragments (with some overlap in the neck). However Pomorski 
and Dietrich~\cite{Pomorski_Dietrich} did the calculation 
in the case of two spherical nuclei and showed that the resulting 
potential is similar to the folded Yukawa-plus-exponential (YPE) 
potential proposed by Krappe, Nix and Sierk~\cite{Krappe_Nix_Sierk}. 
For this reason I use the latter potential, with the parameters 
of Ref.~\cite{Sierk_rotating_nuclei}.

The actual calculation of the different contributions of TKE for the
three paths requires to determine a scission configuration for each path.
As for the symmetric CF valley, I postulate 
that the shape at $\Dcm=15$~fm in Fig.~\ref{shapesFm258CFsym} 
is the scission-point configuration. This is consistent with 
the criterion of Goutte \textit{et al.}~\cite{Goutte_U238} that, 
at scission, the nuclear density in the neck, at $z=z_{\rm neck}$ 
defined in Sect.~2, is 0.01 $\rm fm^{-3}$. In the CF valley this
corresponds to $\Dcm\approx 15.0$~fm. 
In fact it is remarkable that a scission point 
lies in the bottom of a valley. In contrast, no scission 
configurations are found along the symmetric or the asymmetric 
EF paths. It becomes more difficult and ambiguous to assign a 
scission point to each of these EF paths. However, in order to obtain an 
estimate of the associated kinetic energies, a kind of ``sudden 
approximation'' is used for an approximate calculation of 
$E_{\rm C}^{(\rm int)}$. The two nascent fragments are approximated 
at the exit point by the equivalent coaxial spheroids having the 
same elongations and root-mean-square radii as the actual
fragments. They result from a sudden neck rupture assumed to preserve the mass 
and charge asymmetries as well as the center-of-mass distance. The Coulomb 
interaction energy is calculated from the exact analytical 
expression of Quentin~\cite{Quentin_Coulomb_interaction}, whereas the 
nuclear interaction energy is calculated with the YPE potential by
numerical integration.

The calculated results of $E_{\rm C}^{(\rm int)}$, 
$E_{\rm N}^{(\rm int)}$, and $E_{\rm K}^{(\rm sc)}$ for each path are 
given in Tab.~\ref{scission_Fm258}, together with the resulting
kinetic-energy values rounded to the nearest integer. 
\begin{table}[h]
\caption{Characteristics of the approximate scission configurations 
for each fission path: center-of-mass distance $\Dcm$ in fm, 
corresponding total elongation $Q_{20}$ in barns, deformation parameter 
$\beta$ (dimensionless) defined by Zhao \textit{et al.}~\cite{Zhao_PRL1999}, 
Coulomb interaction energy $E_{\rm C}^{(\rm int)}$, nuclear interaction 
energy $E_{\rm N}^{(\rm int)}$ and prescission kinetic energy 
$E_{\rm K}^{(\rm sc)}$ in MeV, together with their sum TKE and the 
most-probable heavy-fragment mass $A_H$.\label{scission_Fm258}}
\begin{center}
\begin{tabular}{ccccccccccccc}
\hline
\hline
 Valley & $\Dcm$ & $Q_{20}$ & $\beta$ & $E_{\rm C}^{(\rm int)}$ 
& $E_{\rm N}^{(\rm int)}$ & $E_{\rm K}^{(\rm sc)}$ & TKE & $A_H$ \\
\hline
 sym. CF  & 15.0  & 289.26 & 1.27 & 238.6 & $-$1.7 & 7.1 & 244 & 129 \\
 asym. EF & 18.2  & 462.14 & 1.55 & 204.7 & $-$1.4 & 11.7 & 215 & 141 \\
 sym. EF  & 22.0  & 714.36 & 1.86 & 178.5 & $-$5.2 & 10.4 & 184 & 129 \\
\hline
\hline
\end{tabular}
\end{center}
\end{table}
The TKE-values for the symmetric CF and EF valleys lie in the ranges 
of the experimental high- and low-energy modes, respectively (see Fig.~7 of 
Ref.~\cite{Hulet_Fm258_PRC1989}). As for the asymmetric EF path, the HFBCS 
total kinetic energy is less than 220~MeV, which is consistent 
with the conclusion from Fig.~8 of Ref.~\cite{Hulet_Fm258_PRC1989} 
that the fragments associated with the fission events above $A_H=140$ 
in the mass distribution have a TKE-value lower than 220~MeV. The authors of 
Ref.~\cite{Hulet_Fm258_PRC1989} also reported that the mass-yield curve 
obtained by selecting spontaneous-fission events 
with ${\rm TKE}<200$~MeV becomes asymmetric, 
from which it can be inferred that a significant number of 
symmetric pairs of fragments have a TKE-value between 200 and 220~MeV. However 
there is no evidence showing which type of symmetric configurations 
(elongated or compact) dominates in this kinetic energy range, or what
the average TKE-value of fragment pairs with $A_H\geqslant140$~is. 
Therefore, based on kinetic-energy considerations, it seems possible 
that the asymmetric EF valley contributes to feed the low-energy mode, 
but probably less so than the symmetric EF one because of the ridge 
separating them.

Two recent studies support this suggestion. On the one hand, Zhao 
\textit{et al.}~\cite{Zhao_PRL1999,Zhao_PRC2000} deduced a deformation 
parameter of the scission configurations $\beta$, which gives a
measure of the deviation from two touching spheres, from experimental 
average total-kinetic-energy systematics. The value $\beta=1.49$
that these authors obtained from the average TKE-value of the low-energy mode in 
$^{258}$Fm is close to the average value $\beta_{\rm asym}=
1.53\pm0.02$ corresponding to the asymmetric mode throughout the actinide 
region. It is interesting to note that the value I obtained 
for the asymmetric EF path $\beta_{\rm HFBCS}=1.55$ is compatible with this 
systematics. On the other hand, Asano 
\textit{et al.}~\cite{Asano} performed more recently dynamical 
calculations of the fragment kinetic-energy and mass distributions 
of $^{264}$Fm as well as the fragment-mass distributions of 
$^{256}$Fm and $^{258}$Fm, at an excitation energy of the compound 
nucleus of 10~MeV. These authors found that their distributions 
can be decomposed into three 
modes: a mass symmetric high-energy mode ($\overline{\rm TKE}=232.1$~MeV), a 
mass asymmetric, low-energy mode ($\overline{A_H}=147.0$, 
$\overline{\rm TKE}=200.8$~MeV) and a symmetric, very low-energy mode 
($\overline{\rm TKE}=171.7$~MeV). Their calculated deformation parameters of 
the scission configurations associated with the first two modes are in very 
good agreement with the systematics of Zhao \textit{et al.}\ 
~\cite{Zhao_PRL1999,Zhao_PRC2000}. Despite the results obtained by
Asano and collaborators correspond to a 
different isotope at a higher compound-nucleus excitation energy, they can 
be considered similar to those for $^{258}$Fm reported in 
Tab.~\ref{scission_Fm258}.

Contrary to the above interpretation proposed for the calculated fission
paths, Warda and collaborators~\cite{Warda2002,Warda2003} identified the 
low-kinetic-energy mode with the asymmetric EF path only, since they do not 
seem to have found a symmetric 
EF path. In the same way as for $^{256}$Fm, they accounted for the 
symmetric character of the corresponding mass 
distribution by a mass symmetric division associated with left-right
reflection asymmetric shapes. This explanation is not supported 
by the present HFBCS calculations. However, it would be very interesting to 
compare the exit points of the asymmetric EF valleys obtained in both
models. Indeed the finite-range effect of the Gogny effective 
force may play a role in the nuclear interaction energy, therefore
impacting the scission configurations and the total kinetic energies.

Finally the similar abundance experimentally observed for the two fission 
modes still needs to be explained. Since a static model cannot 
predict the branching ratio, I can only provide the following plausible 
qualitative argument. As we can see in Fig.~\ref{cutPESFm258}, 
the symmetric EF path appears between $\Dcm=12.0$~fm 
and $\Dcm=12.5$~fm, at about the same energy as the CF path. 
This seems to indicate an equally important feeding of both valleys, 
leading to an expected branching ratio of about~1.

\section{Conclusion}
\label{conclusion}

The features of the potential-energy surface of the $^{256}$Fm and $^{258}$Fm 
isotopes calculated within the HF(SkM*)+BCS(G) model successfully 
account for the experimentally observed asymmetric-to-symmetric 
transition in the mass distribution for spontaneous fission as 
well as most of the measured properties of the bimodal spontaneous fission 
in $^{258}$Fm. The HFBCS results suggest a different interpretation from 
the one proposed by Warda \textit{et al.}~\cite{Warda2002} for the 
low-energy mode in the spontaneous fission of $^{258}$Fm. This mode seems 
to be better understood as a combination of a dominant one 
corresponding to symmetric very elongated scission configurations, 
and another one associated with asymmetric rather elongated
scission shapes (with a nearly spherical heavy fragment). 
The estimates of the total kinetic energy for each component are compatible 
with the experimental data~\cite{Hulet_Fm258_PRC1989} as well as 
with recent dynamical calculations~\cite{Asano}. As an alternate 
dynamical approach, the virial-theorem-based approach, developped 
and initially applied to heavy-ion collisions by I. N. Mikhailov 
and collaborators~\cite{virial_fusion}, can also be applied to fission. 
This work is underway for the symmetric fission of the $^{258}$Fm isotope 
with microscopically calculated ingredients, namely the potential-%
energy surface obtained in the present study and the inertia parameters 
calculated in the HFBCS framework as in Ref.~\cite{mass_parameters}.\\

\section*{ACKNOWLEDGEMENTS}

I am deeply grateful to Arnie Sierk and Peter M\"oller for a careful 
reading of the manuscript and helpful comments, and I would like to
thank Philippe Quentin and Patrick Talou for valuable discussions. 
This work has been supported by the U.S. Department of Energy under 
contract W-7405-ENG-36.


\end{document}